\documentclass[12pt]{article}
\usepackage{graphicx}
\usepackage{amssymb}
\usepackage{epstopdf}
\title{Matter and Light in Flatland}
\author{Omar Y\'epez\\
Department of Chemistry, Memorial University of Newfoundland, \\
St. JohnÕs, NF, A1B 3X7
Canada.}
\begin{document}
\maketitle
\begin{abstract}
In order to explain self-interference phenomenon, a new particle model, based on the intersection of a higher dimensional object with a lower dimensional world, as the source of classical and quantum properties, is proposed. This model suggests that particles are made with a non-material signed current, which travels at the speed of light, through three new curved dimensions. These dimensions occur at atomic scale-lengths (10$^{-15}$ m) rather than at the Planck length (10$^{-33}$ m). Thanks to this, the old problem of equal sign walls huge electric repulsion force, in the electric sphere model is solved, since these curved dimensions confine these walls, preventing them from coming apart. Moreover, this model also suggests that real fermions are not 3D spheres but 4D toroidal particles, which, by their own nature, print a sinusoidal electric field pattern that occupies all the dimensions of the 3D space when moving (matter wave), and takes a toroidal shape at rest. According to the model, flat fermions present two non-superimposable mirror images (enantiomers) with different 2D magnetic dipole moment orientations. And these 2D magnetic dipole moments appear as the temporal intersection of its 3D toroidal dipole moments (anapole moment). As a consequence, a separation of these enantiomers occurs under a magnetic field in flatland, which is consistent with the Stern-Gerlach results. Since femtometer toroidal structures have been found in nuclei, at its ground state (Mz = 0), this model is consistent with the probable shape of real fermions at such state. As a bonus, the model also suggests the probable structure of real photon, which is a 4D toroid also, but contrarily to real fermions, this object prints a sinusoidal electric field pattern that occupies one dimension less from the three available in 3D space when moving (electromagnetic wave) and it will not rest. As a consequence of the comparison between how the 2D space is intersected by flat fermions and flat photon, mass has been identified as the way in which the lower dimensional world is intersected: Òif the higher dimensional object intersection uses all the dimensions of the lower dimensional world, the object presents mass, whereas if that intersection uses one dimension less from the dimensions available in the lower dimensional world, which is the case of flat photon, the object presents no massÓ. Self-interference and the number of turns before being identical are explained for flat fermions and flat photon in 2D space, placing in the same footing those two rather different objects. Probable Fermions geometric properties as well as densities and the magnetic dipole moments are also deduced. Correlations experiments are described for flat electrons, therefore, uncertainty principle appears as a consequence of stochastic interchange between different ways of intersecting the plane by flat fermions and it is an intrinsic space-particle property. Finally, as the number of very different phenomena can be explained with the same model, Òthe intersection of a higher dimensional object in a lower dimensional world is proven to be a powerful concept to explain matter and lightÓ.
\end{abstract}

\section{Introduction}
At this moment, and with the most conservative proposal, string theory is applying ten dimensions to explain matter and gravity. Since space-time uses four dimensions, the move of theoretical physic has been to reduce the other six to a tiny and undetectable dimensional-ball, at the size of the Planck length (10$^{-33}$ m)\cite{kaku}. However, those extra dimensions can be used as the source of a real fermion and photon model, because it presents a way to understand how the huge electric repulsion force that should occur at those tiny distances (10$^{-15}$ m) does not throw apart the fermion in pieces.
On the other hand, "particle-wave" duality could be understood if the particle has more dimensions than the space where it is. By this way, the higher dimensional "particle" leaves a "wave" pattern on the lower dimensional space (vacuum) as it travels through it. The deep relation between matter and space has been demonstrated through general theory of relativity, even though, that theory does not explain why matter bends the space to produce the gravitational field.
The higher dimensionality proposed\cite{kaku}, would be very productive if fermions and photon use, in any way, those extra dimensions to exist and interact with the space. Nonetheless all the efforts and due to a lack of a model, which could explain wave and particleÕs behavior; how are real Fermions and real Photon, what produces mass and how many dimensions constitute each object, are open or not even formulated questions! For instance, it is believed that mass is due to the presence of a Higg boson \cite{gribbin177}, which impairs mass to objects which need it. However, impairing the property of mass will not explain why mass bends the space. All these problems will have a solution if one finds the internal structure and dimensionality of quantum objects and their relationship with the vacuum.\\
One of the most significant experiments, which strongly suggest that quantum objects have more than three dimensions, is self-interference. Single photon interference experiment is a well established but, also an unexplained phenomenon. In describing it, Feynman wondered, Òthere must be some sneaky way that the photon divides in two and then comes back together againÓ \cite{feynman}. In this paper, it will be shown how this sneaky way could be. He also said that this experiment encapsulate Ôthe central mysteryÕ of quantum mechanics. It is Ôa phenomenon which is impossible, absolutely impossible, to explain in any classical way, and which has in it, the heart of quantum mechanics. In reality, it contains the only mystery...the basic peculiarities of all quantum mechanics. Therefore, the explanation of self-interference is crucial in the understanding of this theory.\\
Since self-interference is a common phenomenon for photons\cite{zou, awaya}, electrons \cite{zou, gribbin112}, neutrons \cite{awaya} and even atoms \cite{gribbin112}. Its explanation is of a paramount importance to understand the quantum world. For instance, if one assumes that the quantum particle could be in two places at the same time, this quantum particle should have, at least, more than three dimensions. This feature introduces the possibility that a quantum particle may have a complex structure in space-time. Assuming that such structure is possible, one can arrive to the shape of a higher dimensional winding spiral current twisted toroid, because this object will have construction and destruction components in the same body, and therefore its intersection with a lower dimensional world would explain self-interference, as will be described.
The results presented in this paper suggest that real fermion should be a 4 D torus, which has two ways of intersecting 3D space: 1) in which the object intersects 3D space in two different places at the same time, producing a waving pattern and 2) a toroidal structure when at rest. Recently\cite{forest}, femtometer toroidal structures have been detected in nuclei at rest, which corroborates the model, presented here.\\
In this paper, a particle-space model that describes the quantum behavior for flat fermions (2D electron, proton and neutron) and flat photon (2D photon) is derived. The same visualizes at least 5 different dimensions for flat quantum particles in flatland (2D space) and contains the following features:

\begin{enumerate}
  \item An intrinsic phase shift, which produces construction and destruction electric field vectors upon intersection with flatland, which explains self-interference.
  \item Different symmetries, which explains the number of turns before being identical for flat quantum particles.
  \item Leaves a wavelike electric field in flatland when the object travels through it.
  \item The probable origin of inertia and mass.
  \item The appearance of two equivalent flat fermions enantiomers with different magnetic dipole moment orientations, which undergoes a separation while traveling in an external magnetic field.
  \item The probable shape of a real fermion as a toroidal winding current, the appearance of a toroidal dipole moment (anapole moment) and its time dependant (intersection with 2D space) detection as a magnetic dipole moment in flatland. As well as real Fermion radius, volume, density and magnetic dipole moment values.
  \item The probable origin of uncertainty principle as an intrinsic property of the quantum object.
  \end{enumerate}

\section{Model}
The quantum world is strange, because it involves a more than three dimensions problem. One is constrained to look at the intersection of a higher dimensional object in a three-dimensional world, therefore one is very limited to imagine how this higher dimensional object could be, because is only able to observe its intersection with 3D space. In order to probe in that world, one just has to observe the higher dimensional object itself, and deduces what features of its intersection with the space coincides with the experimental observations. To accomplish this, in the following figures, one component of our familiar three-dimensions is reduced as minimum as possible, therefore our familiar three dimensional vacuum become a paper sheet, a plane, i.e. flatland (2D space). Under these conditions, our detectable third dimension is actually the fourth one (time). In this 2D space, matter and light particles (which will be described) are symmetrically embedded and leave in the plane what one can measure.\\
To make the flat fermion model, take two metal slinkies, place pieces of masking tape on each metal spire of each slinky, in a way that each piece of masking tape is going in a straight line through the slinky. Following the line of masking tape pieces, trace a mark with ink on one side of all the masking tape marks already done; it will produce arrows. Since the final geometric object one is constructing is a toroid, the internal toroidÕs angle (which is in the metal spire or circle) is defined as $\theta$ and the external one (which is around the toroidÕs equator) as $\phi$. Take one slinky and twist it a full turn (360$^{\circ}$) clockwise in the $\theta$ direction. Do the same with the other, but in the counterclockwise direction. Join the spiral head mark with the spiral tail one in each slinky, this produces two twisted toroids. Place the clockwise toroid already made on the right hand and the counterclockwise on the left hand. On Fig. 1 it is shown the result of this procedure. As one moves these toroids like their arrows indicate, it is observed that in the  $\phi$ axis, the toroids go in a relative counter direction.\\
To make the flat photon model, do same procedure as before but with a double  $\theta$ turn (720$^{\circ}$). On Fig. 2 the result is shown, as one moves these toroids like their arrows indicate, it is observed that in the $\phi$ axis, both toroids go in the same direction.

\begin{figure}
\begin{center}
\includegraphics[height=7.2in]{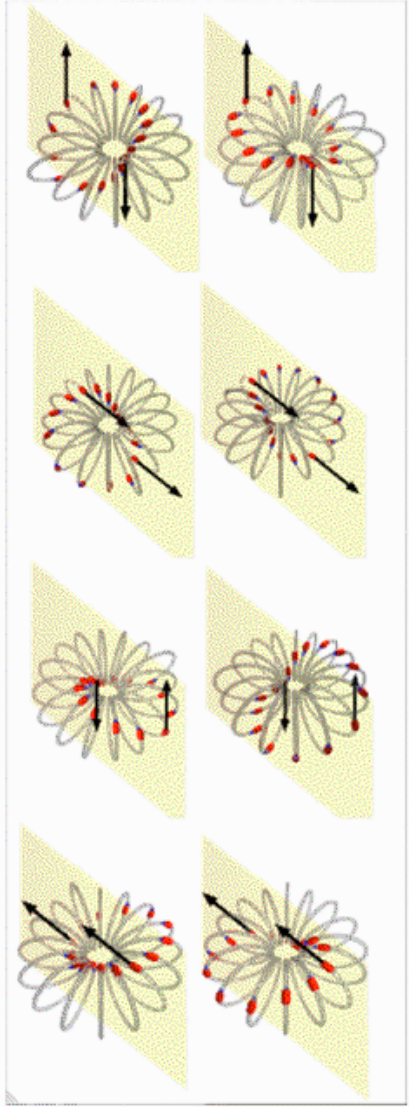}

\caption{Flat fermion intersection with 2D space (flatland) sequence, propagation is downward the page.}
\end{center}
\end{figure}

\begin{figure}
\begin{center}
\includegraphics[height=7.2in]{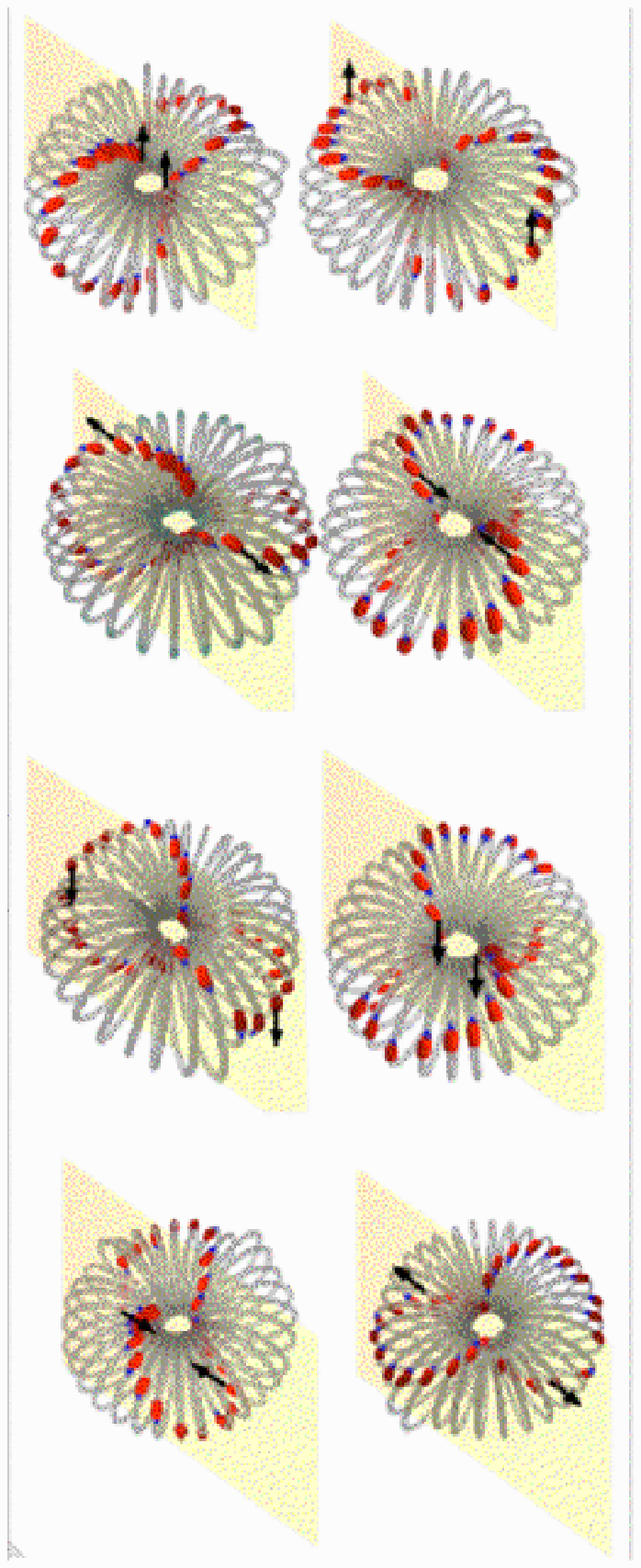}
\caption{Flat photon intersection with 2D space (flatland) sequence, propagation is downward the page.}
\end{center}
\end{figure}

\section{Structure description and interaction with flatland}
Several spatial dimensions form the twisted toroidal structure. One of those dimensions runs along the slinkies' spiral ($\theta$ axis), which contains a non-material signed current that travels at the speed of light, \textit{c}. Another one is a spiral dimension that produces the turns in the $\theta$ axis (one full turn for flat fermions and two for flat photon). The third extra dimension detected is where the slinky's spiral makes the full turn in the $\phi$ axis. All these plus the two dimensions of flatland makes a total of 5 dimensions for flat quantum particles.
On Fig. 1 it is seen the sequence of intersections with the plane, performed by the flat fermion, as it travels through it (downward the page). This sequence begins with a zero electric field state, followed by a maximum, another zero state and a minimum (maximum and minimum assignment is arbitrary). As it can be noticed, the plane symmetrically divides both toroids, through the $\theta$ axis, leaving half of it structure up and the other half down it. Taking this third dimension as time, the toroids parts up the plane are in its immediate future and the toroidÕs parts down the plane are in its immediate past, i.e. the toroid has a complex 2D space-time structure referring to flatland.
In this figure also, one can see that all toroids located at the left are the mirror images (using the plane as a mirror) of the right ones. Flat fermion $\phi$ rotation is deeply associated with $\theta$Õs. Each $\phi$ = 90$^{\circ}$ produces a $\theta$ displacement equal to 90$^{\circ}$. When this object travels through the plane, they do it with a rotation in the $\phi$ axis and a correspondent rotation in the $\theta$ axis as well. Their current spirals cross the plane in and out with a helix movement, at a certain intersection speed, V$_{\phi}$, and at four places in that space. This intersection process will spend some time; therefore, it will leave four time-dependant current vectors in the plane, i.e. electric field vectors. The phase shift in the $\theta$ axis makes that these four electric field vectors have different orientations, which produces a resultant after vector summing. This resultant change in magnitude and direction as the toroids travel, finally producing a sinusoidal electric field printed in the plane, SEFPP (see Fig. 3 and 4). 
\begin{figure}
\begin{center}
\includegraphics[height=7in]{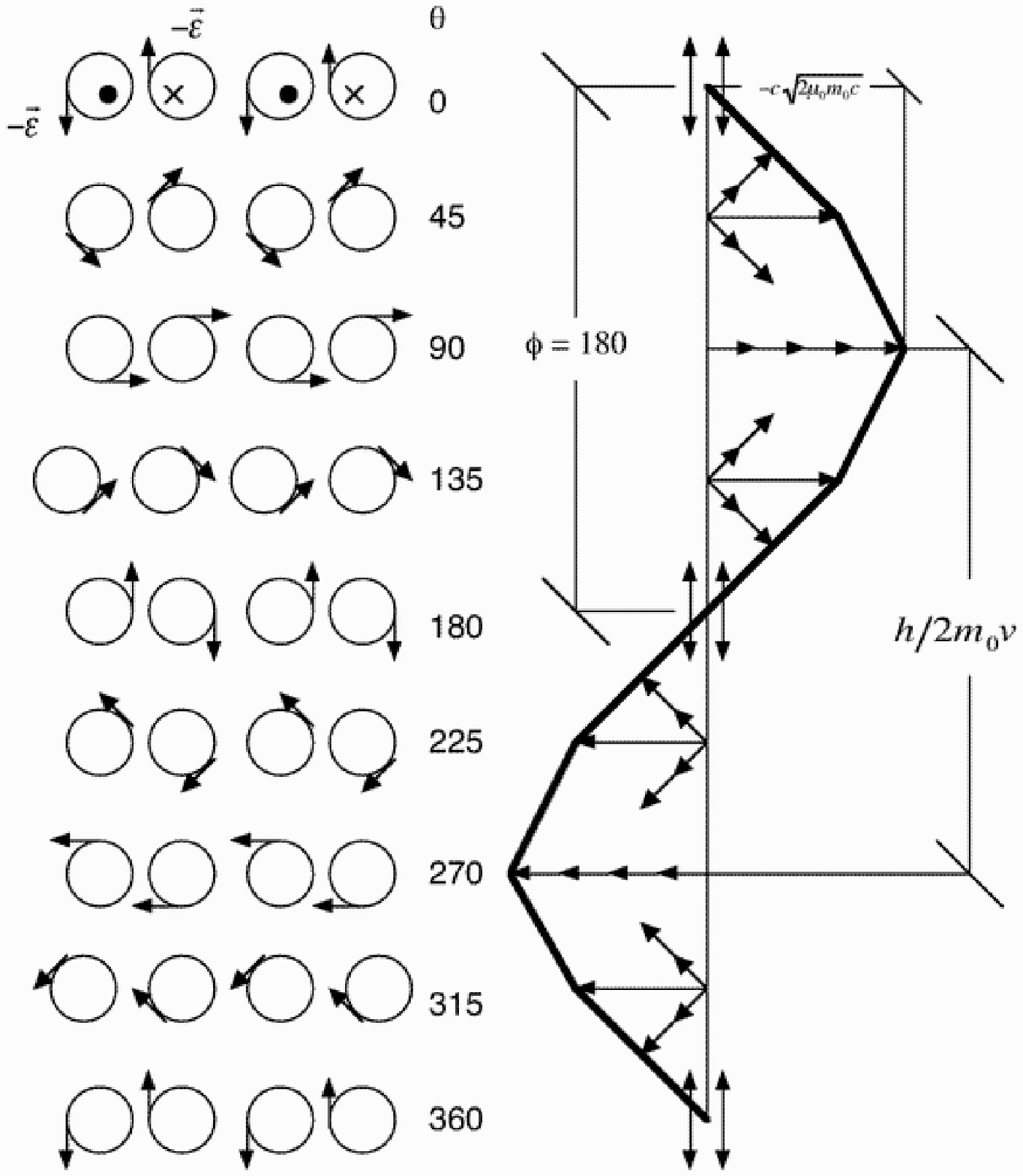}
\caption{Sinusoidal Electric Filed Printed in the Plane by flat fermion. Maximum and minimum go perpendicular to the direction of movement, which is downward the page.}
\end{center}
\end{figure}
\begin{figure}
\begin{center}
\includegraphics[height=7in]{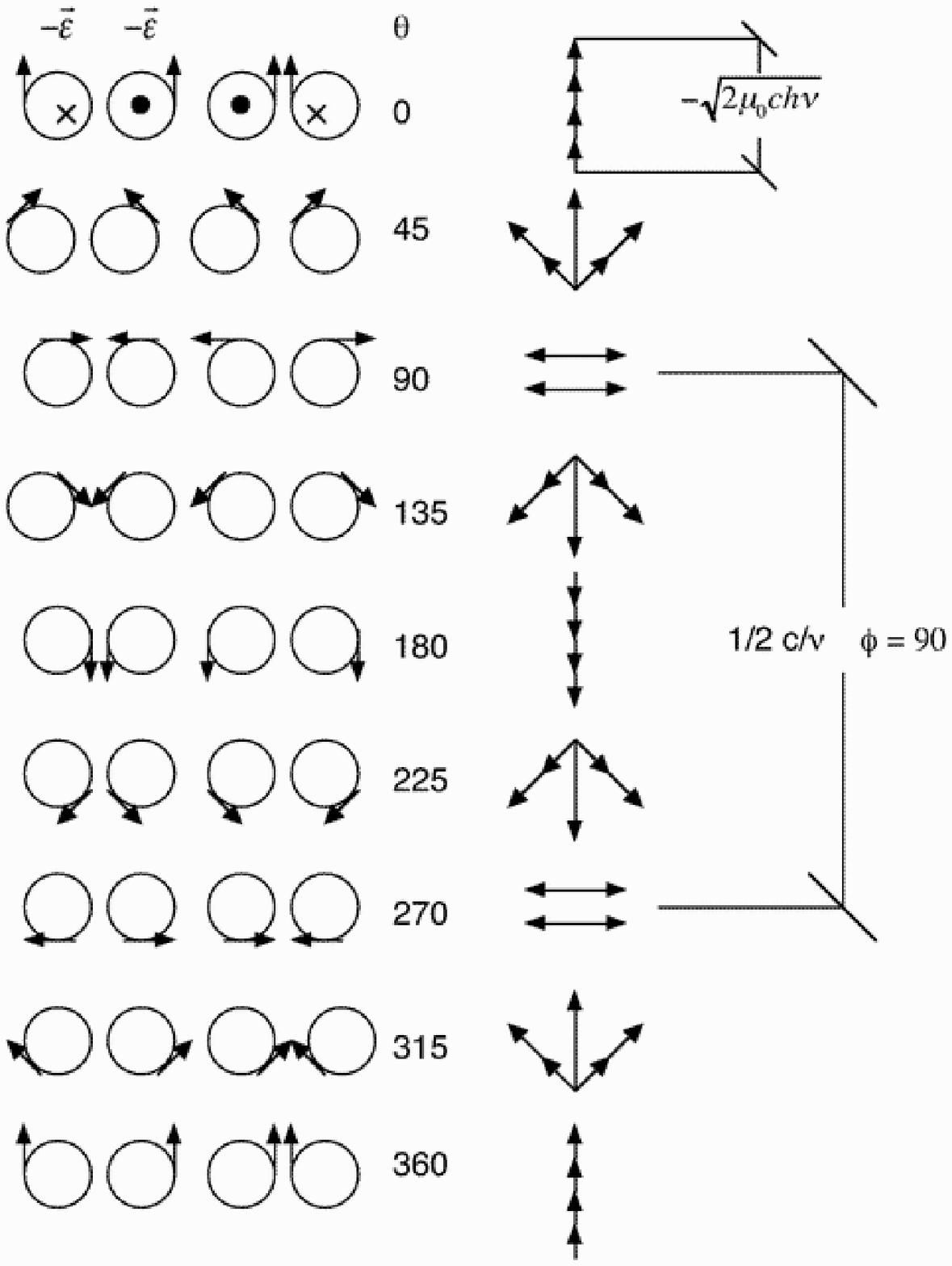}
\caption{Sinusoidal Electric Field Printed in the Plane by flat photon. Maximum and minimum go in the direction of movement, which is downward the page.}
\end{center}
\end{figure}
As it is clear from this explanation, the plane is where the measurement is done and therefore, it is at t = 0 (present). For clarity purposes, both toroids are shown separated although they occupy the same place. On Fig. 1 it is also observed that, in the $\phi$ axis, flat fermionÕs toroids rotate one against the other. As a consequence of this, its movement through the plane will depend on external forces. In this sense, the wavelength of the SEFPP made by this object depends solely on its traveling speed through the plane (see Fig. 3). The $\phi$ rotational speed, V$_{\phi}$ is the same as the traveling speed, in order to leave the electric field vectors in the plane on time. Finally, another consequence of this structure is that the $\theta$ perimeter length is independent of the flat fermionÕs SEFPP wavelength, which is the De Broglie wavelength\cite{eisberg79},
\begin{equation}
\label{debroglie}
\lambda=\frac{h}{m_{0}v}
\end{equation}
Where $\lambda$ is the object wavelength, \textit{h} is the Plank constant, \textit{m$_{0}$} is the inertial mass and \textit{v} is the speed of the object (see Fig. 1 and 3). As it can be noticed, this SEFPP is described by the particle wave function.
On Fig. 2 the flat photon is presented, it begins with a maximum, followed by a zero electric field vector state; a minimum and another zero state (downward the page). The same features as in the case of flat fermion are detected. However, all toroids located at the left are oriented at $\phi = 90^{\circ}$ with respect to the right ones. As a consequence of this structure, there are symmetry relations between given states, i.e. the maximum is equivalent to its minimum after a C$_{2}$ perpendicular to the plane (180$^{\circ}$ edge) and a mirror image (using the plane as a mirror, see Fig. 2). The same applies for both zero states. This means that they are equivalents, i.e. undistinguishables. In this case also, $\phi$ rotation is also deeply associated with $\theta$Õs. However, due to symmetry, flat photon produces undistinguishable states each $\phi = 90^{\circ}$, having two equivalents maximums and minimums in a full $\phi$ round. Due to its symmetry, each $\phi = 180^{\circ}$ occurs a complete $\theta$ round. Since V$_{\phi}$ = \textit{c}, V$_{\theta}$ = 0.5 \textit{c} in order to leave the electric field vectors in the plane on time. In this case, both toroids rotate in the same direction in the $\phi$ axis, therefore the object is not resistant to be moved and its movement does not depend on external forces. As it can be inferred, inertia is a consequence of the flat fermion internal structure and it is a property that flat photon does not have.
On Fig. 3 and Fig. 4, the SEFPP for flat fermion and flat photon can be observed respectively. It is clear from these figures that flat fermionÕs SEFPP looks like a transversal wave, whereas flat photonÕs looks like a longitudinal one.

\section{Self-Interference}
The proposed structures for flat fermions and flat photon easily explain why a single flat fermion or a single flat photon could arrive as a maximum or as a minimum, depending on the angle measured from the interference slits. The already known requirement that the distance between slits should be an integer number of the particle wavelength, $\lambda$; to observe an interference pattern, just satisfy an internal structure requirement, which is that the quantum object has to be in a maximum state just at the interference slits. Since upon intersection with the space, flat fermion and flat photon produces construction and destruction electric field vector states and that just depends on its internal structures, they will be able to produce a maximum or a minimum upon arrival to the detector screen. This will only depends on the optical path for both branches of the toroid intersection. Therefore, the maximums and minimums of a single quantum particle will follow the same equations for maximums and minimums already known respectively\cite{holiday}, 
\begin{equation}
\label{maximo}
d\sin\alpha=p\lambda
\end{equation}
\begin{equation}
\label{minimo}
d\sin\alpha=(p+\frac{1}{2})\lambda
\end{equation}
Where \textit{d} is the separation between slits, $\alpha$ is the angle of the maximums or minimums measured from the slits, $\lambda$ is the wavelength of the flat photon and/or flat fermion (De Broglie wavelength) and \textit{p} is an integral number. Different angles from the slits will produce different optical paths = different phase interference, thus producing the complete sinusoidal interference pattern observed \cite{zou, awaya, gribbin112}. On Figs. 5-6, the self-interference graphics to produce a minimum and maximum for flat fermions and flat photon respectively are shown. When one of the slit holes is blocked, there is no possibility for the flat fermion and flat photon to be in both slits (two places at the same time) and therefore, there will not be an interference pattern.
\begin{figure}
\begin{center}
\includegraphics[height=7in]{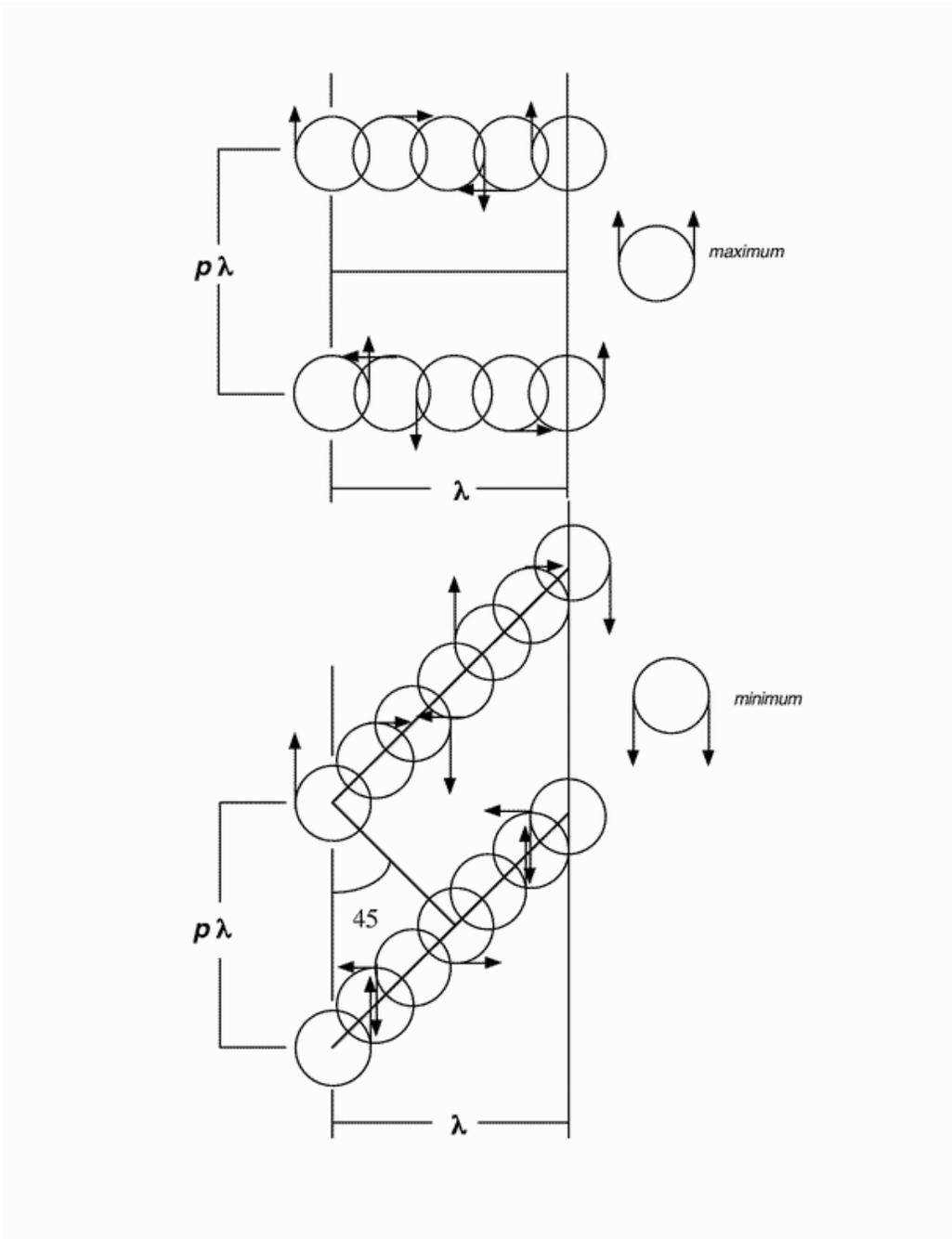}
\caption{Maximum and Minimum in the flat fermions self-interference experiment.}
\end{center}
\end{figure}
\begin{figure}
\begin{center}
\includegraphics[height=7in]{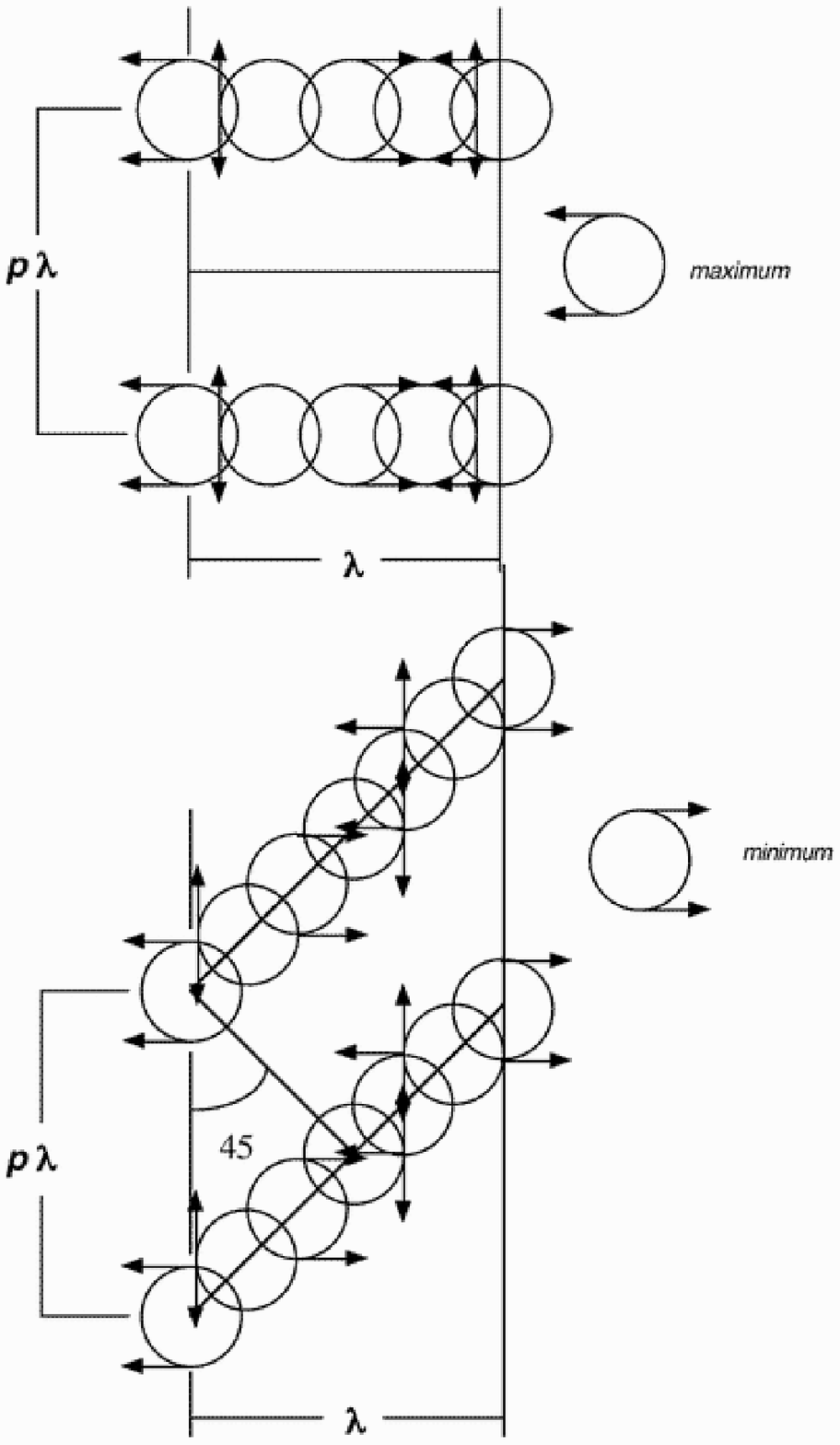}
\caption{Maximum and Minimum in the flat photon self-interference experiment.}
\end{center}
\end{figure}
\section{Number of Rounds before being Identical}
A very intriguing aspect, which differentiates real fermions from photon is that 3D fermions make two rounds before arriving to its initial state, whereas 3D photon just makes one \cite{gribbin371, battey}. As one can see on Fig. 1-2, however, a toroid, not a sphere, is what is doing such rounds. Since the toroid intersects symmetrically the plane through its $\theta$ axis. On Figs. 3 and 4, both branches of the toroid, sums $2\pi r_{\theta}$ radians when they arrive to $\theta=180^{\circ}$, where $r_{\theta}$ is the $\theta$ intersection radius. Thus, half a toroidal $\theta$ round can be interpreted as a particle round. Given this explanation, it is easy to observe that on Fig. 3, there is a round between both zero electric field states ($\theta = 0^{\circ}$ and $180^{\circ}$). However, these two zero electric field states are not equivalent from the point of view of the 3D object (c.f. Fig. 1); as a matter of fact, they are not equivalent with any symmetry operation. A flatlander observes that the object did a complete round, but it is not the same. Then it observes that after another round, the flat fermion arrives to its original state. As a consequence, its complete SEFPP wavelength is made after two nonequivalent and complete rounds.
In the case of flat photon, on Fig. 4, there is a round between its maximum and its minimum ($\theta = 0^{\circ}$ and $180^{\circ}$). However, both states are equivalent after a $C_{2}$ and a mirror image (using the plane as a mirror, see Fig. 2). Therefore, a flatlander observes that after a complete round, the object is the same. As a consequence, the flat photon complete SEFPP wavelength is made after two equivalent and complete rounds.

\section{Flat Fermion Properties}
\subsection{Flat Fermion Radiuses, Volumes and Densities}
As it was described, in the case of flat fermion, there is no relation between its SEFPP wavelength and its $\theta$ perimeter. The flat fermion SEFPP wavelength is related only with its speed (see Fig. 3). To look at the flat fermion $\theta$ perimeter, the quantum particle should be collided by a high-energy flat photon (2D X ray photon) in a frontal collision (180$^{\circ}$)\cite{eisberg59}. By this way, the Compton shift of the dispersed 2D Xray photon will be equal to the $\theta$ perimeter, i.e.
\begin{equation}
\label{compton}
\lambda_{C_{180^{\circ}}}=\frac{2h}{m_{0}c}
\end{equation}
This wavelength is the measure of the flat fermion $\theta$ perimeter. By using this distance, the $\theta$ circle radius for the flat fermion toroid would be,
\begin{equation}
2\pi r=\lambda_{C_{180^{\circ}}}=>r_{\theta}=\frac{h}{\pi m_{0}c}
\end{equation}
Assuming that the flat fermion toroid, at rest, is a horn torus, which seems reasonable from ref. 1, its area will be,
\begin{equation}
S=\frac{4h^{2}}{m_{0}^{2} c^{2}}
\end{equation}
its volume,
\begin{equation}
V=\frac{2h^{3}}{\pi m_{0}^{3} c^{3}}
\end{equation}
and the density D will be,
\begin{equation}
D=\frac{\pi m_{0}^{4} c^{3}}{2h^{3}}
\end{equation}
the radius, diameter, toroid surface, volume and densities of real fermions are shown on Table I.
\begin{table}
\begin{center}
\includegraphics[height=2.0in]{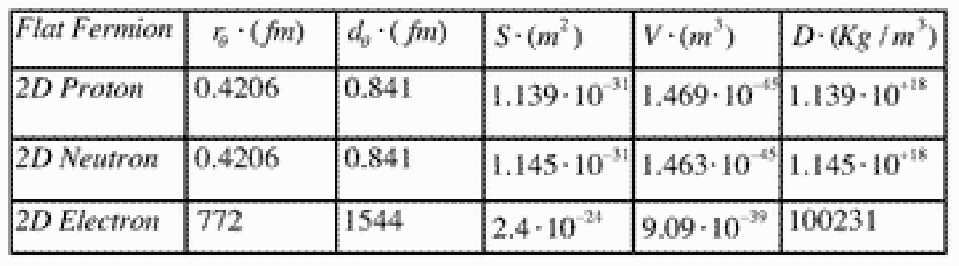}
\caption{Flat particles geometric data}
\end{center}
\end{table}

\subsection{Flat Fermion Magnetic Dipole Moment and Charge}
From Fig. 1, it can be observed that flat fermions are sets of 3D winding current toroids, which are in the same place. Because of this arrangement, these toroids present a 3D toroidal dipole moment (anapole moment)\cite{dubovic}. However, as described before, the toroid intersection with the plane spends some time; thus, its 3D anapole moment would be detected in the plane as a 2D magnetic dipole moment. In the following paragraphs, the derivation of such 3D anapole moment and its temporal projection in the plane, to produce a magnetic dipole moment are developed,
Assuming that the fundamental electric charge, e, is uniformly distributed on the surface area of a horn torus in 3D space (see Table I) and if the 2D charge (the charge in the $\theta$ circle) is, in the same way, proportional to the area of the $\theta$ circle, the $\theta$ current, $i_{\theta}$, can be estimated. First, the time to cross the $\theta$ perimeter, $t_{\theta}$, is,
\begin{equation}
c=\frac{\frac{2h}{m_{0}c}}{t_{\theta}}=>t_{\theta}=\frac{2h}{m_{0}c^{2}}
\end{equation}
the 2D charge will be,
\begin{equation}
Q_{2D}=i_{\theta}t_{\theta}
\end{equation}
therefore, the following surface to charge relation can be established,
\begin{equation}
\frac{\frac{4h^{2}}{m^{2}_{0}c^{2}}}{e}=\frac{\frac{h^{2}}{\pi m_{0}^{2}c^{2}}}{\frac{2i_{\theta}h}{m_{0}^{2}c^{2}}}
\end{equation}
thus,
\begin{equation}
i_{\theta}=\frac{em_{0}c^{2}}{8\pi h}
\end{equation}
the number of 3D toroidÕs spires can be obtained multiplying this current by the time needed to have the charge, \textit{e}; converting this time to a distance and dividing this result by the $\theta$ perimeter, i.e.,
\begin{equation}
i_{\theta}=\frac{em_{0}c^{2}}{8\pi h}=\frac{e}{t}=>t=\frac{8\pi h}{m_{0}c^{2}}=>d=c\cdot\frac{8\pi h}{m_{0}c^{2}}=\frac{8\pi h}{m_{0}c}=>n=\frac{\frac{8\pi h}{m_{0}c}}{\frac{2h}{m_{0}c}}=4\pi
\end{equation}
Finally, the 3D space toroidal moment (anapole moment) can be obtained through\cite{dubovic},
\begin{equation}
T=\frac{1}{4\pi c}\cdot IV
\end{equation}
where I is the current in its coil, which is,
\begin{equation}
I=ni_{\theta}
\end{equation}
and V is the toroid volume. By combining (7), (12), (13) and (15) in (14), one has,
\begin{equation}
T=\frac{4\pi}{4\pi c}\cdot \frac{em_{0}c^{2}}{8\pi h}\cdot \frac{2h^{3}}{\pi m_{0}^{2}c^{3}}=(\frac{\hbar}{m_{0}c})^{2}e
\end{equation}
However, as it was discussed, this toroidal dipole moment is made with spiral currents that spend some time intersecting the plane, and producing a 2D time dependant toroidal dipole moment, i.e. a magnetic dipole moment. On Fig. 7, the magnetic dipole moment produced by this process can be observed, if one uses as a 
\begin{figure}
\begin{center}
\includegraphics[height=7in]{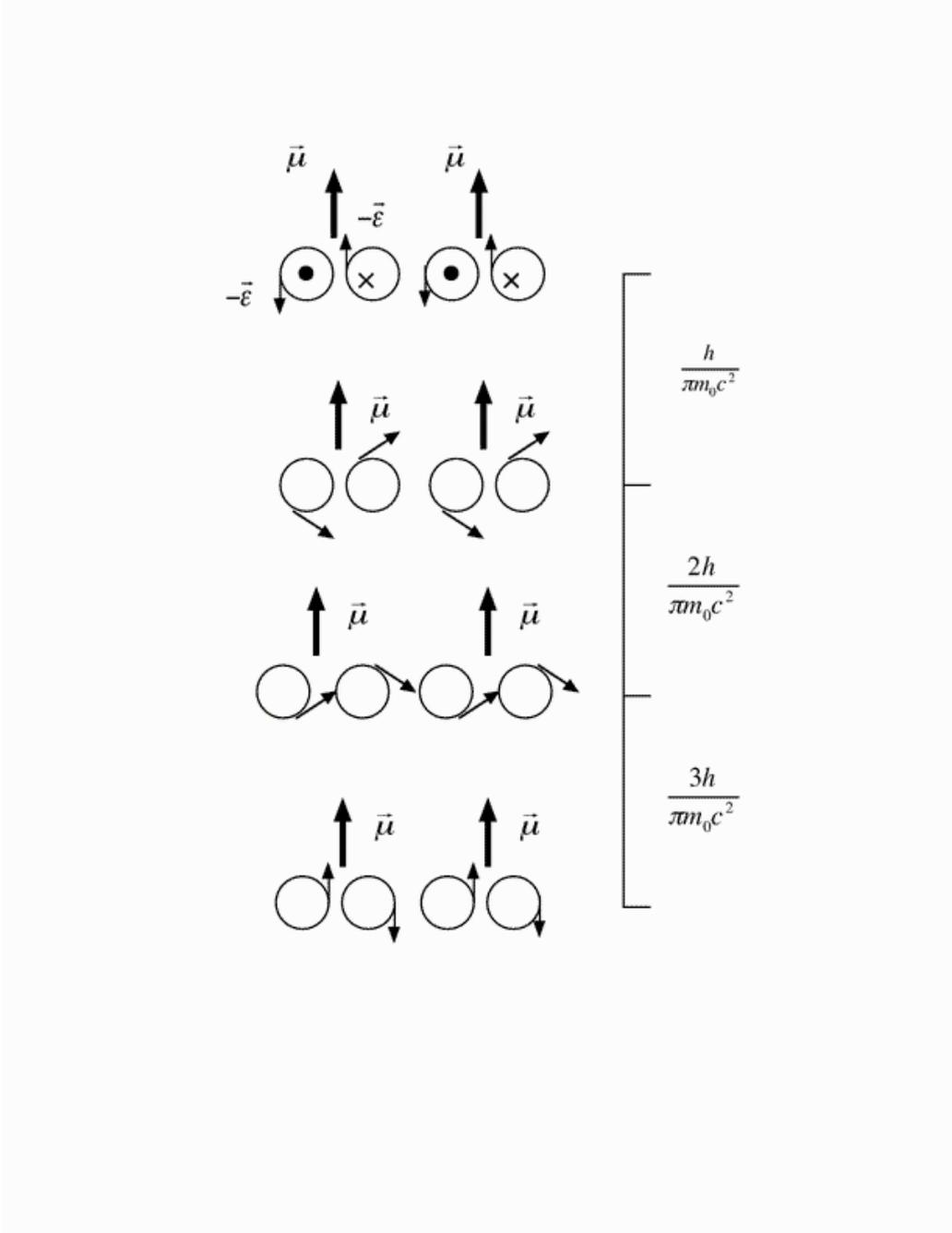}
\caption{Temporal intersection in flatland by the flat fermion 3D anapole moment, producing the magnetic dipole moment, ${\mu}$.}
\end{center}
\end{figure}
intersection time, $\frac{h}{\pi m_{0}c^{2}}$\, as shown in that figure, one obtains for the electron,
\begin{equation}
\mu_{e}=\frac{(\frac{\hbar}{m_{0}c})^{2}e}{\frac{h}{\pi m_{0}c^{2}}}=\frac{\hbar e}{2m_{0}}=9.28\cdot 10^{-24} A\cdot m^{2}
\end{equation}
in the case of the neutron,
\begin{equation}
\mu_{n}=\frac{2(\frac{\hbar}{m_{0}c})^{2}e}{\frac{h}{\pi m_{0}c^{2}}}=2\frac{\hbar e}{2m_{0}}=9.66\cdot 10^{-27} A\cdot m^{2}
\end{equation}
In the case of proton, however, the value still gives 2.79 $\hbar e/2m_{0}$.\\
The 2D electric charge would be related to the current, $i_{\theta}$, multiply by the period of time which this current spends to travel the $\theta$ perimeter and by two toroid intersections, this will be,
\begin{equation}
2\cdot Q_{2D}=2\cdot \frac{em_{0}c^{2}}{8\pi h}\cdot \frac{2h}{m_{0}c^{2}}=\frac{e}{2\pi}
\end{equation}
As could be observed on Fig. 8, this $\theta$ current should have a sign, in order to produce a signed 
\begin{figure}
\begin{center}
\includegraphics[height=6in]{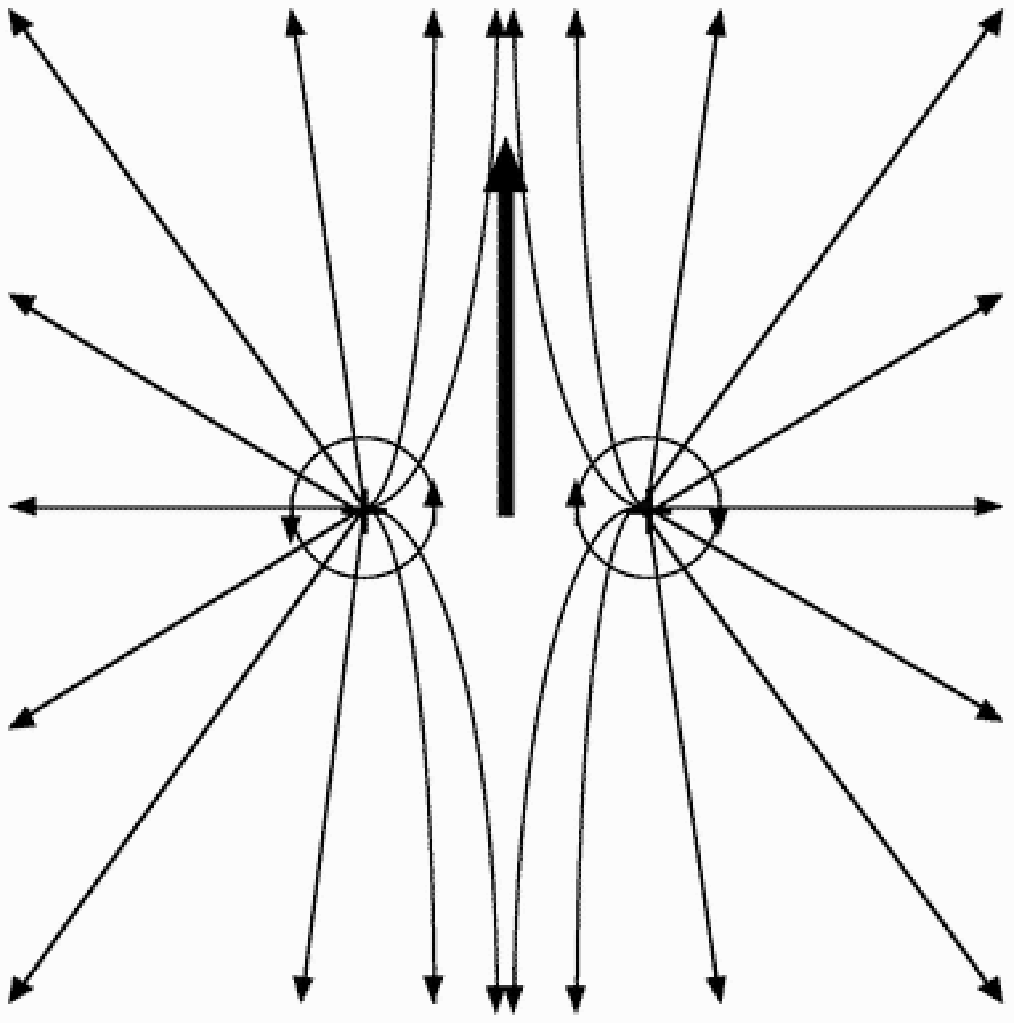}
\caption{Electric Field produced by positive signed current flat fermion, the magnetic dipole moment is also shown between both circles.}
\end{center}
\end{figure}
charge and thus an electric field in flatland, as shown. And this has no connection with the direction of the current in the spires. On the other hand, in the case of flat photon, it should have opposite sign currents between both toroids, in order to have no electric field outside the particle, as it is expected.
\subsection{Matter and Light Wave Amplitude}
As it was stated in the previous section, the spiral current spends some time intersecting the plane. This process leaves electric field vectors, which magnitude can be derived as follow,
\begin{equation}
\overrightarrow{\varepsilon}=-\frac{d\overrightarrow{E}}{dr}=-\frac{Rd\overrightarrow{i_{\theta}}}{cdt}
\end{equation}
Where \textit{R} is the spiral resistance in the distance and or time $dr=cdt$ and $di_{\theta}$ is the $\theta$ perimeter current in the period of time, $\frac{\lambda}{\pi c}$ (for flat photon) or $\frac{h}{\pi m_{0}c}$ (for flat fermion), which is 1/6 of the spiral current $i_{\theta}$. Taking into account that the intensity of the wave is related with the wave electric field amplitude through \cite{eisberg89},
\begin{equation}
In=Nh\nu=\frac{1}{\mu_{0}c}\frac{\overline{\varepsilon}^{2}_{m}}{2}
\end{equation}
Where \textit{In} is the wave intensity, $N$ is the number of particles per square meter per second, which in this case is 1 $m^{-2}s^{-1}$, $\mu_{0}$ is the vacuum magnetic permeability and $\varepsilon_{m}$ is the wave electric field amplitude. Also, according to the model, regardless being a flat fermion or a flat photon, the sum of four electric field vectors, $\overrightarrow{\varepsilon}$, is the wave electric field amplitude, $\varepsilon_{m} $. Therefore, the spiral resistance and the wave electric field amplitude for the flat photon and the flat fermion are, respectively,
\begin{equation}
R(\Omega)=\frac{12}{e}\sqrt{2N\mu_{0}h\lambda^{3}}=>\varepsilon_{m}(V/m)=- \sqrt{2N\mu_{0}ch\nu}
\end{equation}
\begin{equation}
R(\Omega)=\frac{12}{e}\left[\frac{h}{m_{0}c}\right]^{2}\sqrt{2N\mu_{0}m_{0}c}=>\varepsilon_{m}(V/m)=-c\sqrt{2N\mu_{0}m_{0}c}
\end{equation}
\section{Parity, Stern-Gerlach result and Correlation experiment}
On Fig. 9\textit{a}, the cross section of a flat fermion toroid and its enantiomer (mirror image = parity 
\begin{figure}
\begin{center}
\includegraphics[height=7in]{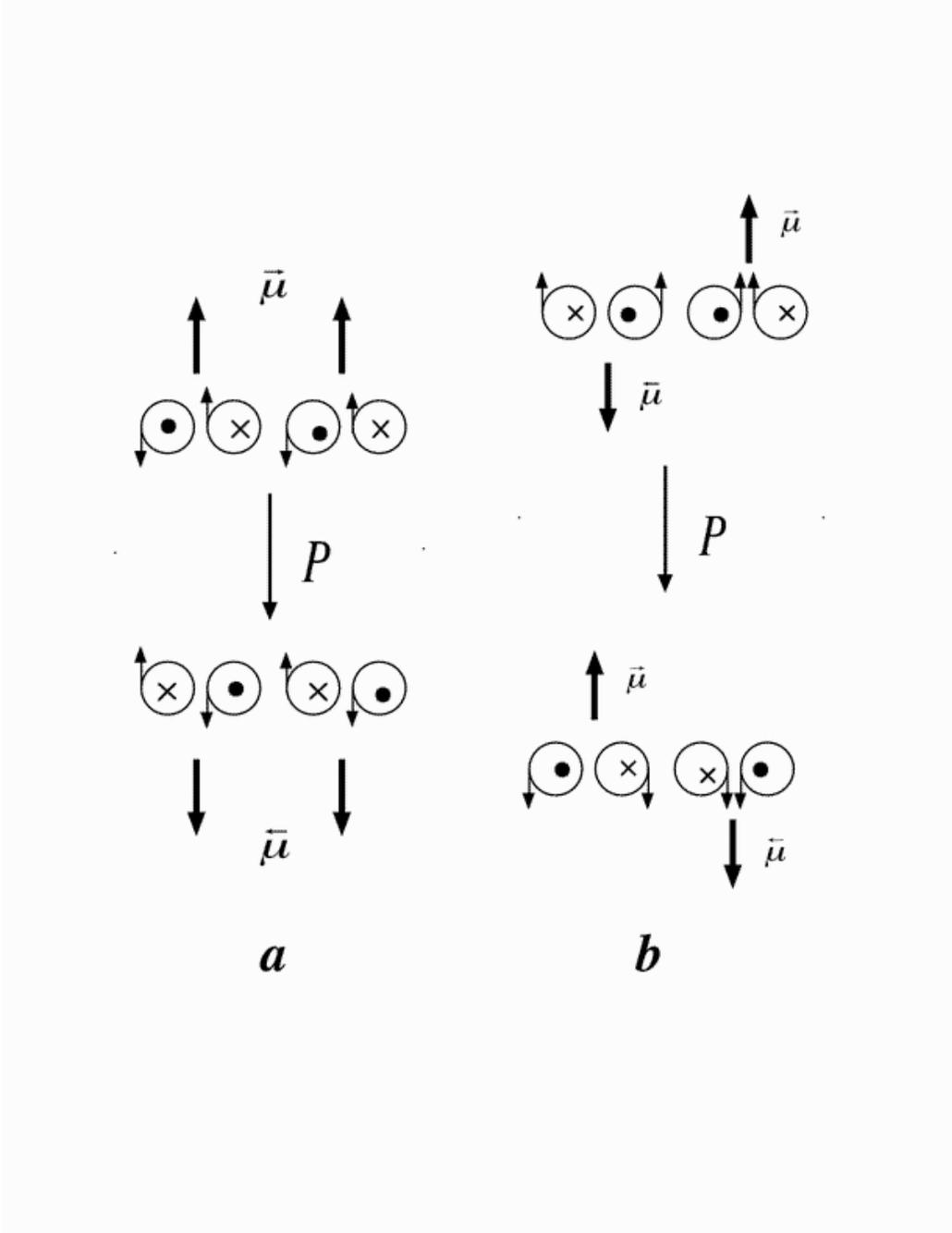}
\caption{Parity transformation of a flat fermion \textit{a} and flat photon \textit{b}.}
\end{center}
\end{figure}
transformation, \textit{P}) produced with a mirror perpendicular to the plane\footnote{Actually, this can be appreciated in Fig. 1 because the right one toroid is the mirror image of the left one, just the direction of the current in one of those have to be changed, to produce a mirror image with a mirror perpendicular to the plane.} and the effect on the orientation of its magnetic dipole moment are shown. It is clear from this figure that, there are two non-superimposable enantiomers; notice that the current vectors will not fit after a $C_{2}\,(180^{\circ}$ edge perpendicular to the plane) and, as a consequence, a change in the orientation of its magnetic dipole moment is produced. Also, in Fig. 9\textit{b}, the parity transformation for the flat photon is shown. It is clear that the flat photon has an internal zero magnetic dipole moment, and contrarily to flat fermion, this particle is invariant after a P transformation (all the current vectors fits after the $C_{2}$ operation).\\
On Fig. 10, two flat fermion enantiomers are traveling through an external magnetic field \textit{B}, which is perpendicular and coming out of the plane (depicted by the dots). This external magnetic field interacts with the magnetic dipole moment (and/or the internal magnetic field)
\begin{figure}
\begin{center}
\includegraphics[height=7in]{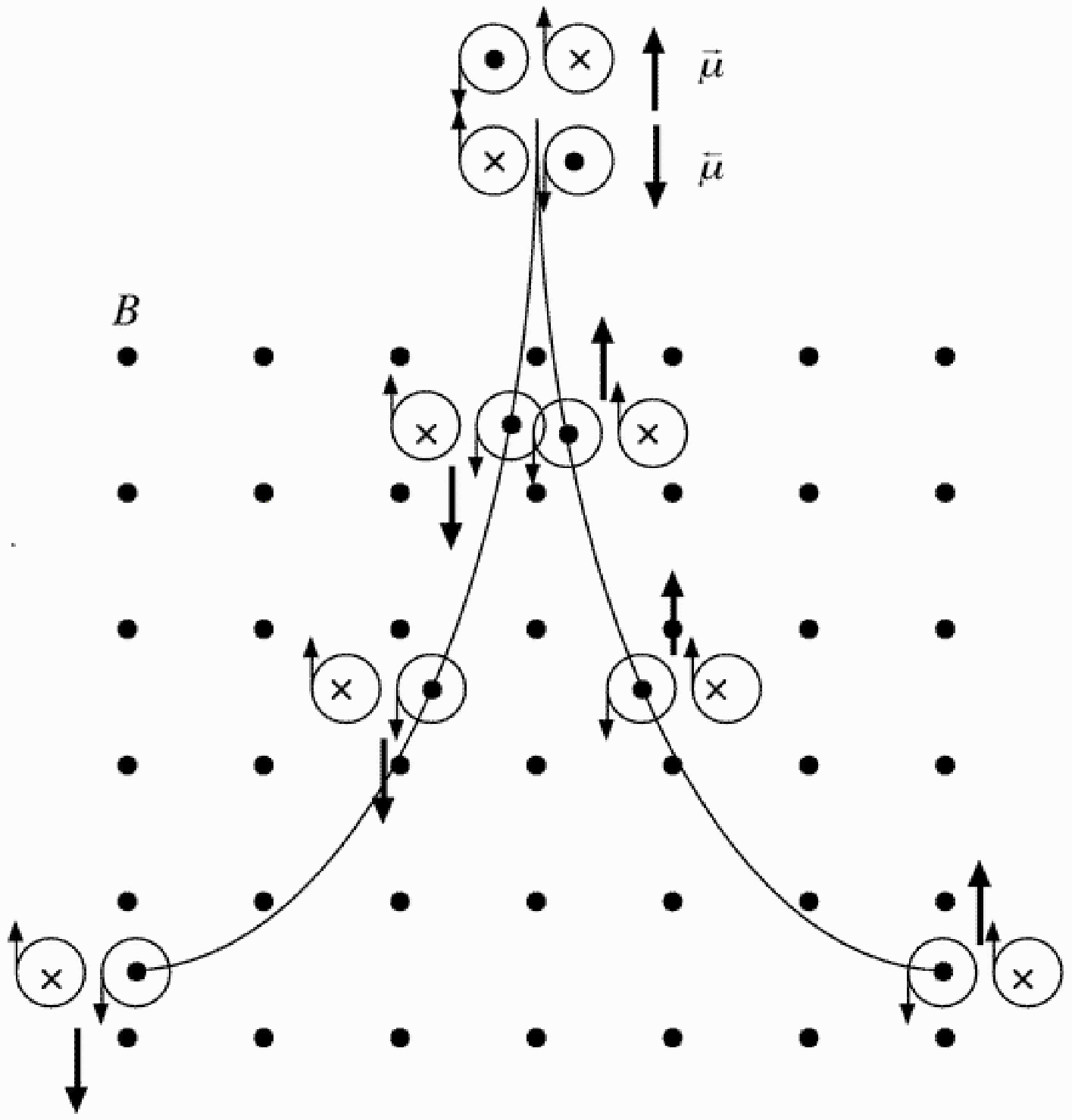}
\caption{Stern-Gerlach experiment in flatland.}
\end{center}
\end{figure}
in each enantiomer separating them as shown (use vectors product  where the first vector is the magnetic dipole moment, $\overrightarrow{\mu}$). This was checked experimentally with little magnets arranged and passed through an external magnetic field as shown in the figure. Finally, a separation to right and left occurs, which is consistent with the Stern-Gerlach results. In the case of flat photon, Fig. 9\textit{b} shows that that object will not produce such results, since this object have a zero magnetic dipole moment.\\ On Fig. 11, 32 electrons pass through a magnetic filter, which consisted in blocking the electrons deviated to the right, the other leftist ones passed through such filter (right filter). Later, the orientation of the external magnetic field is  changed to $90^{\circ}$ to the left. Due to the change in the external magnetic field orientation, the electron intersection with the plane changes also, from the $\theta$ like intersection, shown in all previous figures, to a $\phi$ like intersection. This is a 50:50 process, thus half of the electron magnetic dipole moments aim above and perpendicular to the plane and the other half aim in the counter direction. In this form, the electrons are retarded or accelerated through this particular magnetic field, but non-is blocked because the blocking device is out of the plane. As a result of this process, 8 electrons passed through this particular filter. Upon going back to a $\theta$ like orientation, a new rearrangement of the $\phi$ electrons occur, again by chance, dividing the filtered electrons in the two possible orientation $\theta$ electrons can have. Thus, from the 8 electrons that passed the $\phi$ oriented magnetic filter, 4 are blocked to the left and 4 are passed to the right. Finally a 50 $^{\circ}/_{\circ}$ correlation appears very clearly, each filter divides the amount of electrons in two, but a particular magnetic field orientation produces the elimination of the previous filtering process, because switching between $\theta$ and $\phi$ like electron intersection with the plane. Also, it can be appreciated that the electrons which passes the right filter 
\begin{figure}
\begin{center}
\includegraphics[height=8in]{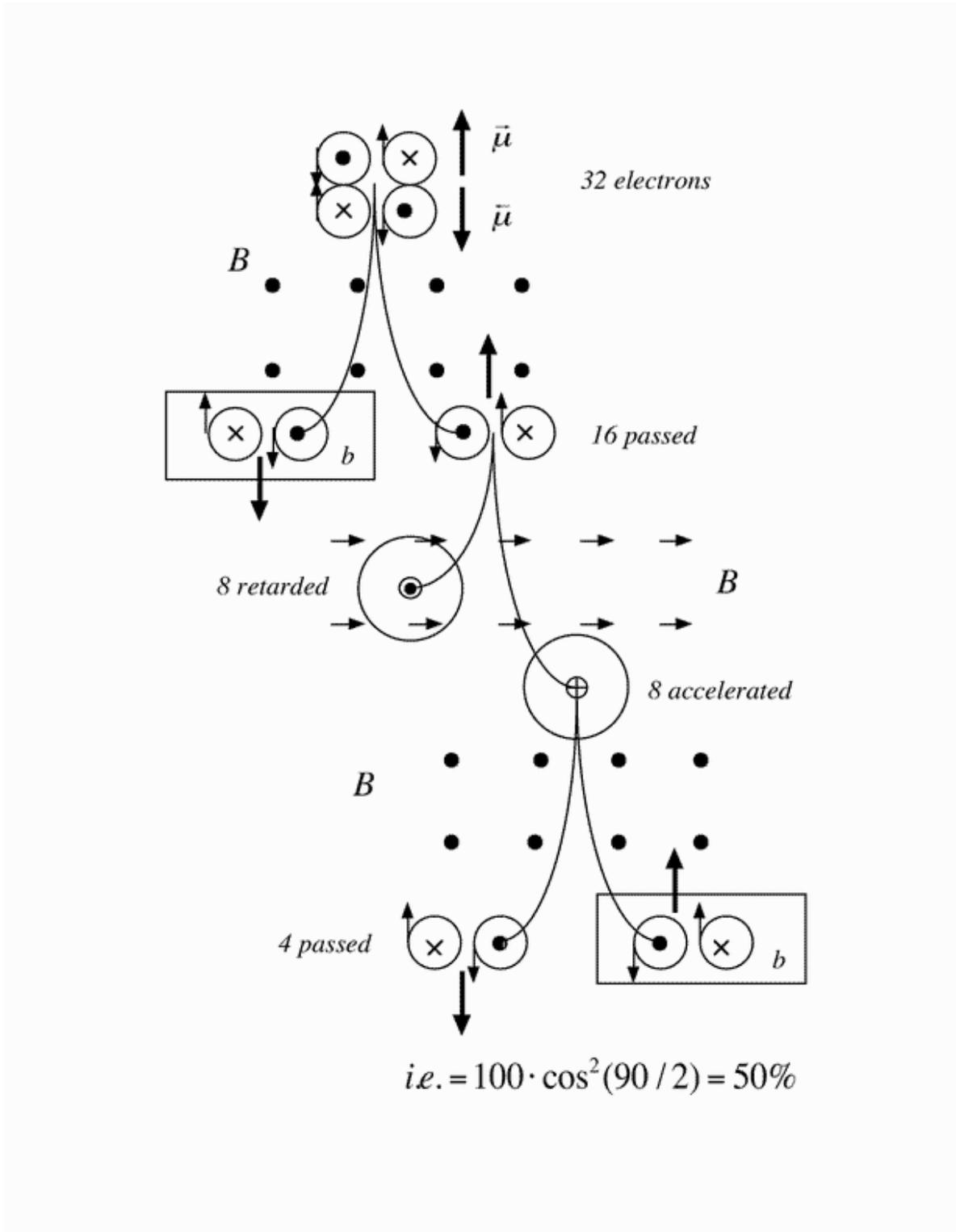}
\caption{Flat electrons through magnetic filters with different orientations}
\end{center}
\end{figure}
will be stopped by a left filter (0 $^{\circ}/_{\circ}$ correlation) and that they will pass another right filter (100 $^{\circ}/_{\circ}$ correlation).\\ 
On Fig. 12 and Fig. 13, Einstein-Podolski-Rosen (EPR) \cite{feynmanIII, gribbin127} like correlation experiments with flat electrons can be observed. 
\begin{figure}
\begin{center}
\includegraphics[height=8in]{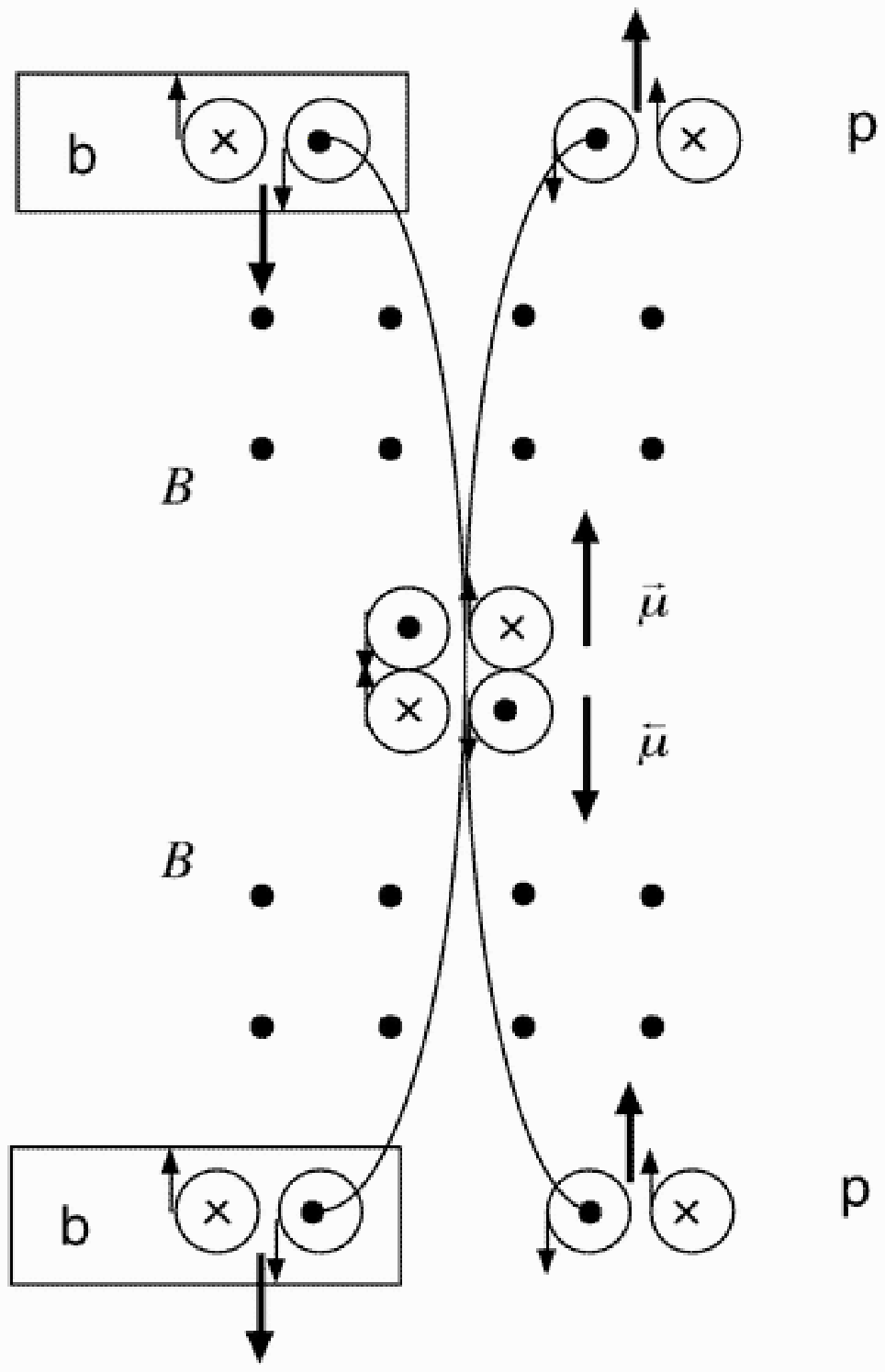}
\caption{EPR experiment with zero correlation}
\end{center}
\end{figure}
\begin{figure}
\begin{center}
\includegraphics[height=8in]{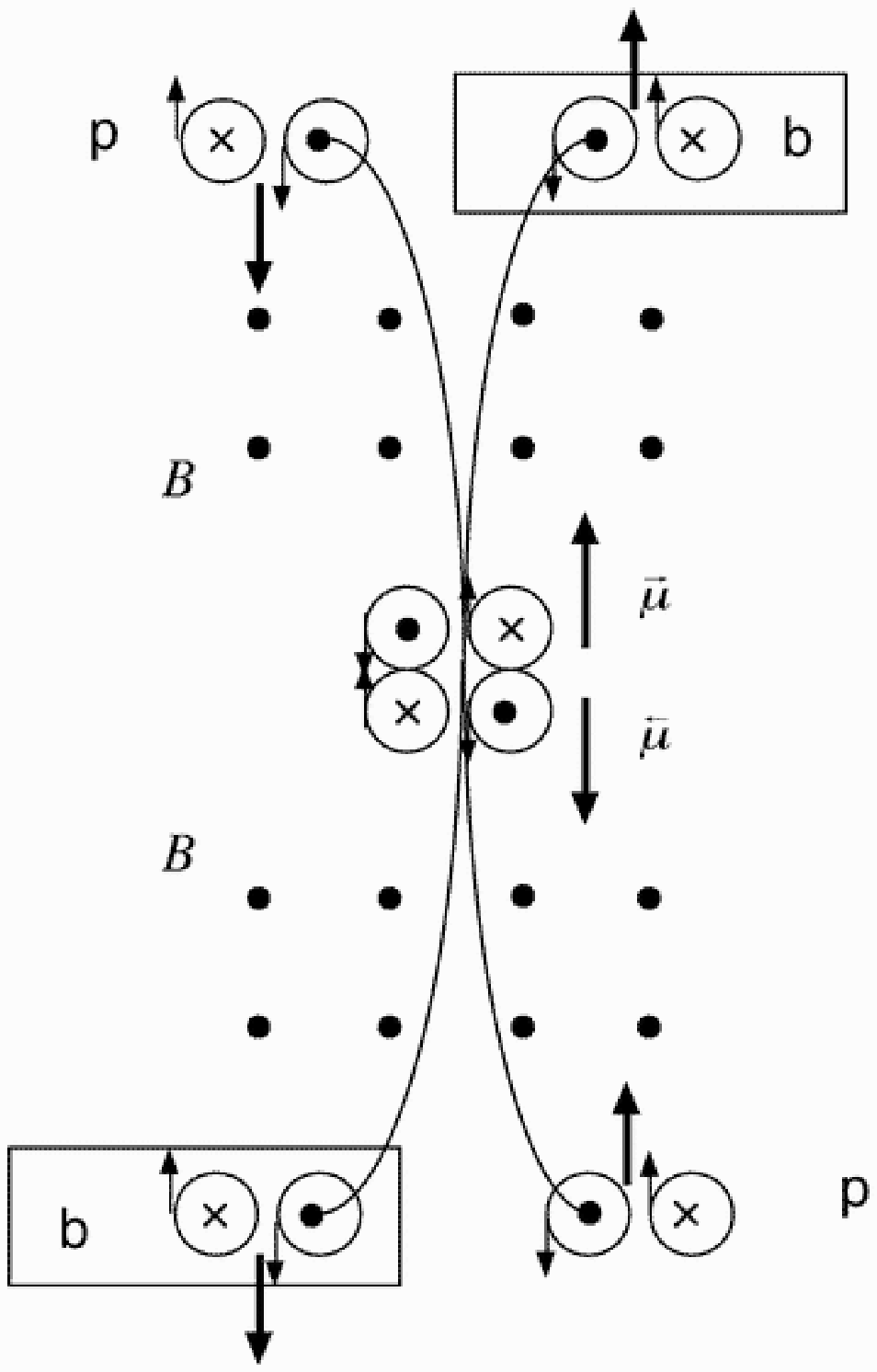}
\caption{EPR experiment with 100 $^{\circ}/_{\circ}$ correlation}
\end{center}
\end{figure}
It is believed that Fig. 12 is the case for equal oriented filters, because both filters allow the passing of the same kind of magnetic dipole moment orientation in the flat electron. It is very easy to realize that if the up going flat electron does pass through the filter, then its down going companion (which will have, always, an opposite magnetic dipole moment)  does not pass its filter, whereas, if the up going flat electron does not pass through the filter, then its down going companion always emerges from its filter.\\ 
It is believed that Fig.13 is the case for counter oriented filters, because both filters allow the passing of opposite kind of magnetic dipole moment orientation in the flat electron. if a particular up going flat electron passes its filter, then its companion down going always passes its filter, whereas, if the up going  electron does not pass its filter, its companion electron doesn't pass through its filter either.
Fig. 14, is figure 13 but changing the orientation of one of the filters to 90$^{\circ}$. Again, the change in the orientation of a magnetic filter from $\theta$ to $\phi$ like intersection, opens the possibility of the reorientation of the magnetic dipole moments at random. As a consequence, if a particular down going electron passes its filter, one-half of the time its companion up going electron will emerge with the same magnetic dipole moment orientation, one-half of the time with the counter orientation.
\begin{figure}
\begin{center}
\includegraphics[height=8in]{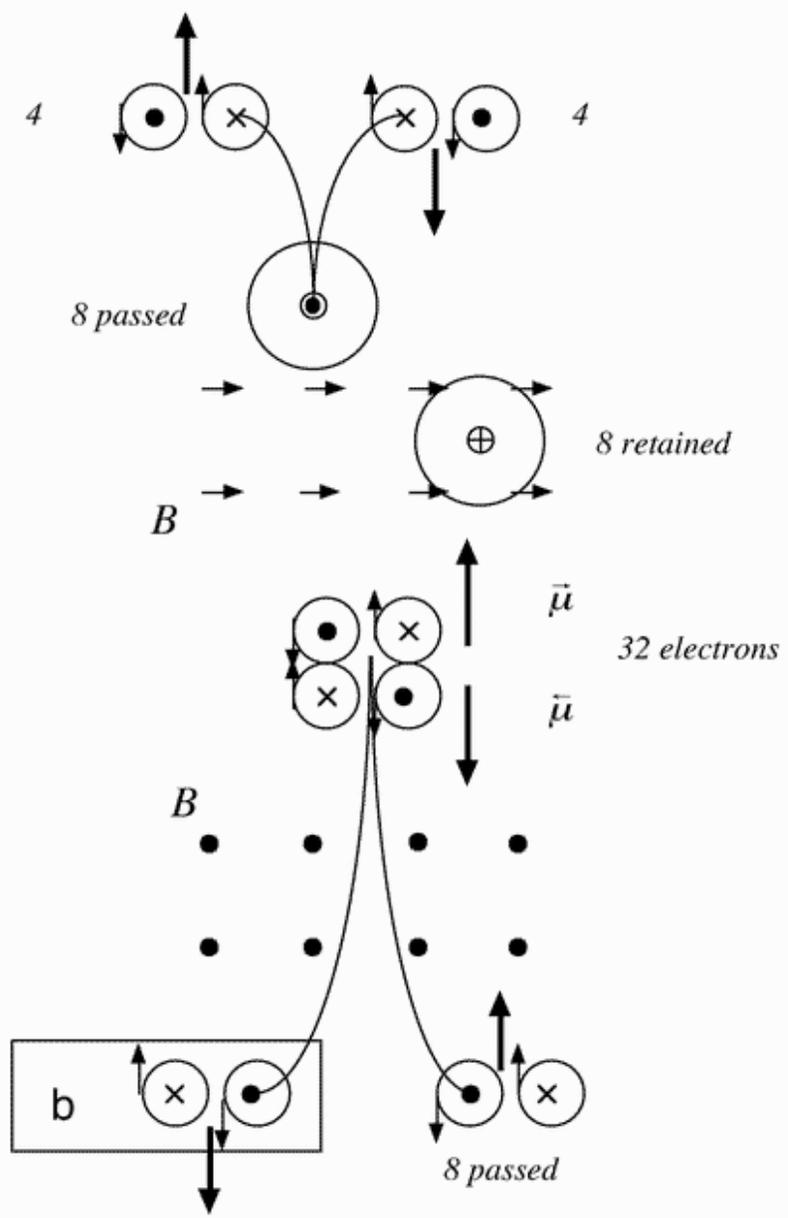}
\caption{EPR experiment with perpendicular magnetic filters}
\end{center}
\end{figure}
\section{Discussion}
\subsection{Flat Fermion}
The new dimensions already described come to solve one of the early mysteries in fermion model development. It was that the charged sphere model should not exist, because the enormous electric repulsion forces, which will occur at those tiny distances, necessarily, will throw apart the sphere's equal sign electric walls, in pieces \cite{feynman28}. This way of thinking drives Dirac (without the actual possibility of higher dimensionality) to propose his point particle. As currents exist in these new curved dimensions, equal sign electric walls built in them, actually exist and they can not go anywhere, because the curved dimensions described confine them. As a matter of fact, femtometer toroidal structures, not sphere ones, has been found in nuclei \cite{forest}, and these structures can only exist, if their charged walls are shaped by those curved dimensions.\\
On the other hand, as the problem of fermion stability is solved with the new dimensions found, and toroidal shapes have been detected for fermions. The interpretation of Stern-Gerlach results, drives one again to imagine a two ways of spinning charged spheres, which will produce two different oriented magnetic dipole moments, depending on the way they spin, and will undergo a separation under an external magnetic field. However, as it was described, the 3D toroidal dipole moment temporal intersection with the plane, produces a magnetic dipole moment also (see Fig. 7), which change its orientation under a parity transformation, producing non-superimposable enantiomers (see Fig. 9\textit{a}), therefore a flatland version of Stern-Gerlach results was found experimentally (see   Fig. 10). By this way, the new model is consistence with Stern-Gerlach results also, with no need to imagine something spinning, but a toroid spending time in intersecting the plane.\\
One advantage of this model is that its hyper-part (the flat particle is in 2D space, its hyper-part is the 3D toroid) is, actually, the 3D intersection of the real model, which would be a 4D toroidal structure. Therefore, one can treat the 3D toroid as the measurable part of the real fermion, which is also corroborated after ref. 1. If this toroid rotates, at rest through its equator edge (an edge through its $\phi$ perimeter), it can be confused with a sphere, which radius will be the toroid internal diameter, two times the $\theta$ radius, $d_{\theta}$, shown on Table I. As a matter of fact, using dispersion-elation fitting, Mainz reports 0.847 \textit{fm}, as the radius of the proton \cite{Karshenboim}, and this value is deviated within 1  $^{\circ}/_{\circ}$ from the theoretical result given in Table I. Hofstadter et al. obtained 0.80 \textit{fm} as the proton radius and concluded that the neutron radius should be very similar to this value \cite{Hofstadter}. This located this neutron radius value at 5 $^{\circ}/_{\circ}$ from the theoretical value given in the same table.\\
Regarding the electron, electron-electron and electron-positron scattering experiments have been correctly predicted by the Dirac theory of the electron, which implies that the electron is a point and/or that its extension is not bigger than 0.001 \textit{fm}. This fact is in direct contradiction with the Lamb shift results, which reveals that the electron's electric charge is smeared out over a region of space that is comparable to the electron Compton radius, i.e. 386 \textit{fm} \cite{enigmatic6}.\\
As described before, when the flat electron is traveling (see Fig. 3). It is spending some time to leave its electric field vectors (this is the time used in equations 17 and 20). The intersection time for a  flat electron is 2.6 $10^{-21}$ s to travel 772 \textit{fm} from a total $\theta$ perimeter of 4852 \textit{fm}, in each $\theta$ circle branch of its toroid. In this interval of time, the flat electron is touching the plane in four places, separated 1544 \textit{fm} at least, to leave four electric field vectors and 1/6 of its $e/2\pi$ 2D charge. Thus, in order to have a collision with the whole flat electron $\theta$ intersection circles, the time spent in the collision, should be 1.5 $10^{-20}$ s or more. If the time spent in a collision is less than this time slot, the collision will happen between point-like structures (sections of the flat electron $\theta$ circles). Thus M\o ller and Bhabba scattering equations (constricted to 2D space) would predict that behavior, which is in agreement with the experimental evidence \cite{enigmatic49}.\\
In the case of the flat proton, the time to completely show its structure is 8.4 $10^{-24}$ s (probably 2.79 times even lower) to travel its 2.6 \textit{fm} of total $\theta$ perimeter. Thus, the flat proton needs 1835 times less time in 1835 times less length, in comparison with the flat electron to show its whole structure in a collision event. 
Just for a comparison, a typical collision time in a nuclear reaction is $\sim 10^{-22}$ s \cite{eisberg676}, which is enough for the flat proton to show up completely, but it would allow just 29 \textit{fm} from the 4852 \textit{fm} of the flat electron $\theta$ perimeter, 0.6 $^{\circ}/_{\circ}$ of its structure, to show up. Increasing the energy of the collision would diminish the time of the collision event, showing a point-like behavior even further. For example, If the collision time is 8.4  $10^{-24}$ s. The flat electron would show only 2.5 \textit{fm}, 0.05 $^{\circ}/_{\circ}$ of its structure. Certainly, increasing the energy of the collision will go against the collision time required for the flat electron to show up. This would be also the case for the flat proton at very high energies, if it would not get destroyed in pieces before showing such behavior. 
Hence a lower energy technique should be used to search another aspects of the flat electron structure. For more or less the same reasons, this has been suggested in other electron model \cite{enigmatic132}.
Coming back to the electron radius, the present model adopts Compton lengths in the toroidal structure, because the flat proton and neutron gave a very accurate result in its radiuses, therefore the flat electron should follow the same trend (eq. 4). Also, the flat electron will have a lower spiral current, $i_{\theta}$ (eq.12) in comparison with the flat proton, thus it has to occupy a bigger space (run a longer time) in order to produce the same charge as the later, and finally being lighter implies that the flat electron should occupy a higher area than the flat proton in the plane, in order to exert a lower tension on it. As it will be discussed, this is the origin of its mass.\\
As one can see, the density values in Table I for the proton and neutron are in the order of magnitude of current estimates $\sim 10^{+18}$ Kg m$^{-3} $ \cite{eisberg595}. Moreover, taking the diameter given in ref. 1, for the deuteron, which is $\sim$ 1 \textit{fm} and using the horn toroid volume (which was used to derivate eq.7), one has
\begin{equation}
V=2\pi^{2}r^{3}=2.5 10^{-45}m^{3}
\end{equation}
The deuteron density is finally 1.36 $10^{+18}$ Kg m$^{-3}$, which is a little bit higher than the proton and or neutron's density reported in Table I, as it would be expected. In the case of the electron, if one uses the corrected Quantum Electrodynamics radius for the electron, which is 670 \textit{fm} \cite{enigmatic5} in equation 24 and using the electron mass, one arrives to a density of 1.5 $10^{5}$ Kg m$^{-3}$, for the electron, which is in good agreement with the result presented in Table I.\\ 
Since the flat fermion 3D toroid has a toroidal dipole moment. The flat fermion 2D magnetic dipole moment was identified as the temporal intersection of this 3D anapole moment, in the plane. By using \textit{e} as the charge of the 3D toroid, the module value of the flat electron and neutron fitted the known values, 1.0012 and 1.91315 $\hbar e/2m_{0}$ \cite{weast} very well (within 1 and 5 $^{\circ}/_{\circ}$ respectively, c.f. equations 17 and 18).
However, the flat proton deviated considerably, being its magnetic dipole moment three times bigger than the experimental value. This could be explained if the intersection time for the flat proton, which is supposed to be $h/\pi m_{0}c^{2}$, is for some reason, 2.79 times shorter than the same for the other two particles (in each particle time scale). Since flat neutron fits the magnetic dipole value, when using 2 \textit{e}, it looks like the flat proton gets a slower intersection time when it traps a flat electron and becomes a flat neutron.\\ 
Charge appears as a quality of the current; therefore positive or negative currents will produce the respective signed charge. By this way an electric field, which is independent from the SEFPP and the magnetic dipole moment, can be produced as shown in Fig. 8.\\
One of the most difficult properties to imagine is a magnetic dipole moment coming from a zero charge object, which is the case of the neutron. However, since the charge is a quality of the current, one can has two same circulating direction but opposite sign toroidal currents, which: will add as the same anapole moment, will produce the magnetic dipole moment upon intersection with the plane, will have the same internal magnetic field orientation and it will only change direction upon parity transformation (see Fig. 9 \textit{a}). Therefore, one can have a magnetic dipole moment built with 2 \textit{e}, but a zero charge object and, this object, still has its enantiomer, i.e. it will produce Stern-Gerlach results.\\
On Fig. 10, the flatland equivalence of the Stern-Gerlach experiment for saying, flat electrons, is observed. Since that universe has only two dimensions, a moving particle can go just  to the right or left, or can be accelerated or retarded. In the case of Fig. 10, a very clear separation between up and down oriented magnetic dipole moments to the left and to the right is observed. This was verified with magnets arranged as described in the figure. Therefore, it is believed that the separation occurs as consequence of the interaction of the internal magnetic field and not the magnetic dipole moment of a given electron with the external magnetic field.\\ 
Fig. 11 is an example of the travel of flat electrons through different oriented magnetic filters in flatland. After passing the first right-filter, the change in orientation of the second filter to 90$^{\circ}$, made the electron to intersect flatland through its $\phi$ axis. This process is at random, therefore, the orientation produced by the first filter is lost. However, the number of electrons is reduced to a half, because in this orientation half of the electrons are accelerated and the other half are retarded (the blocking device is out of the plane). Also, this orientation produces a lost in the identification of the particle because its magnetic dipole moment is out of the plane. Finally, upon another random rearrangement of the 8 accelerated electrons in a left-filter, 4 electrons with their magnetic dipole moment completely opposed to the direction of the magnetic dipole moment of the 16 electrons which passes the first filter, occurs. And this behavior is consistent with the 3D experimental fact \cite{feynmanIII}. \\
The diameter of the $\phi$ oriented electron is $2\lambda/\pi$, its perimeter is $4\lambda$, which is the same as the sum of the perimeters of the two circles in the $\theta$ orientation. If one assumes that the same $\theta$ current occurs in the $\phi$ intersection, the charge is the same in either kind of intersection with the plane.\\
The change in the orientation of the intersection from $\theta$ to $\phi$ like, accompanied with the 50:50 stochastic process is reveling in regard with the uncertainty principle. The $\theta$ orientation is what produces the sinusoidal electric field pattern printed in the plane and it is the way to know the momentum of the particle with precision. However, to do so the particle is in two places of the plane at the same time, therefore, the measurement of its exact position is even nonsensical. It is pure non-local to produce the wave that leaves in the plane. The stochastic change to the $\phi$ intersection stops the particle being non-local, to occupy just one place in flatland, thus its position can be measured very precisely but in doing so, it stops also to produce the sinusoidal pattern, ergo its momentum is completely unknown and there is no way to avoid this, because it is a particle-space intrinsic property.\\
Fig. 12 and 13 show, clearly, that simple symmetry will explain zero and 100 $^{\circ}/_{\circ}$ correlation experiments, respectively. On Fig.14, however, it is clear that knowing the position, i.e. $\phi$ like intersection with the plane of the up-going  accelerated, retained or even stopped electron makes a lost in its identification in 2D space, as well as the information about its momentum, as discussed before. Therefore, the act of measuring the position of the up-going particle, affects the original entanglement and the precision in the measurement of the system total momentum; making it impossible to deduce, with precision, the momentum of the up-going particle by subtracting the momentum of the down-going particle from the system total momentum. Ergo, the EPR paradox is solved without the "spooky action at a distance" \cite{gribbin127}, i.e., \textbf{\textit{uncertainty prevails no matter the arrangement of the experiment, because it accompanies the particle wherever it goes.}}\\
With the increment of one dimension in the model presented, it suggests that the real fermion is a 4D torus, with two ways of intersecting 3D world: one is when it travels and intersects 3D space in two separated places, leaving a sinusoidal electric field that uses all the dimensions of the 3D space and the other when it is at rest, adopting the shape of a torus (in the ground state). It also will have a 4D anapole moment, which temporal intersection with the 3D space will produce a 3D magnetic dipole moment.
Self-interference occurs in all quantum objects and it has been probed even in atoms \cite{gribbin113}. Therefore, different atoms should be a 4D torus in the ground state also. As a matter of fact, femtometer toroidal structures have been detected for $^{2}H, ^{3,4}He, ^{6,7}Li$ and $^{16}O$ \cite{forest}. Although it is beyond the scope of this paper, it is not so difficult to imagine a toroid with all the proton currents alternate with neutron-ones. This line of thoughts would increase the application of this self-interference model to understand nuclear properties.
\subsection{Flat Photon}
As one makes two turns in the $\theta$ axis, all the features observed in the flat fermion's SEFPP change, and another kind of SEFPP appear. In this case, one obtains a sinusoidal pattern also, but it uses one dimension less from the two available in the plane (see Fig. 2 and 4). The double turn in the $\theta$ axis also produces that both toroids move in the same direction in the $\phi$ axis, which was interpreted, as a property of no resistant to be moved by the particle, i.e. no inertia. Contrarily to the case of flat fermion, the double turn in the $\theta$ axis produces that each toroid has opposite orientation in its magnetic dipole moment, therefore, it has an internal zero magnetic dipole moment (see Fig. 9 \textit{b}) and does not produce Stern-Gerlach results, i.e. flat photon will not change the direction of movement under a magnetic field. In this sense, also occurs opposite sign currents between each flat photonÕs toroids, producing a zero external electric field object. Since flat photon is made with two toroids with opposite charges and each toroid has two $\theta$ turns, it is not so difficult to produce two opposite charged flat fermions upon flat photon dissociation \cite{eisberg65}, since all the components for a flat and anti flat fermion are in the flat photon.\\
 Actually, there is no clue about the internal structure of a real photon. The present model indicates the first possible looks to a real photon, albeit it is constricted to a 2D space and it is not a 4D model. It is believed that in Fig. 2 one is looking to a polarized flat photon, because there is no room to allow another plane to wave. In the case of real photon, there will be multiple planes in 3D space and, therefore, one could have photons with different polarization planes. However, in order to observe this; again, one would need to make the intersection of a 4D torus, with the characteristic suggested by the present model, in a 3D space. By this way, it would be possible to observe the real photon sinusoidal current printed in the 3D space picture, which was deduced from MaxwellÕs laws, and which actually occupies two dimensions (its sinusoidal electric and magnetic fields oscillate in perpendicular planes) from the 3 available \cite{alonso}.\\
Given that the flat photon does not change direction in its movement when it pass through a magnetic field and that there is just one plane in flatland, this flat photon model is not suitable to explain its correlation experiments.\\
From the description of the flat fermion and photon structures, the duality wave-particle appears manifest: \textbf{\textit{Fermions and Photon are hyper-dimensional particles, which leaves different kinds of electromagnetic waves in the vacuum}}.

\subsection{Origin of Inertia and Mass}
As flat fermions and photon cross flatland, a sinusoidal electric field is leaved in that world (2D space); the same explains the wave behavior that those Òdifferent particlesÓ show. The De Broglie wavelength is produced as a consequence of the intrinsic resistant to be moved, showed by flat fermions. Also flat fermions use the two coordinates of flatland to travel (see Fig. 3). Flat photon is not resistant to be moved and its SEFPP uses one dimension less from the two available in that space to travel (see Fig. 4). Since real photon has no mass but real fermions do have it: The dimensional condition to have mass is that \textbf{\textit{the object intersection in a given space, should use the same number of dimensions that has such a space}}. As mass occurs when all the dimensions of a given space are used, it exerts a tension on that space, bending it, just as a light metal ring bends the surface of the water when floats on it. Furthermore, if the area of the object is bigger, it will exert less tension on the water surface (2D space), which will tolerate it better. This supports also that the flat electron (less mass but bigger Compton wavelength) is bigger than the flat proton as it has been discussed. In the case of objects that occupy n-1 dimensions from the space they intersect, such tension does not occur and the bending effect is not produced, which is the case of the flat photon. As one can see this concept of mass is consistent with what is proposed by the general theory of relativity \cite{will}, and it is kinetic independent \cite{jammer}.
\subsection{Self-interference and the Interpretation of Quantum Mechanics}
As it is very clear from Fig. 1 and 2, that flat fermion and flat photon are objects, which intersection process with the plane leaves a SEFPP as they travel through it. Therefore, the ad hoc particle-wave duality concept and the principle of complementarity begin to be clarified. Flat fermion and flat photon are higher dimensional particles, which leaves a sinusoidal electric field printed in the plane as they travel through it and therefore, the wave equation applies to describe both kind of particleÕs behavior. Furthermore, since the intersection process leaves this electric field vectors in the plane, flatlander witness the appearing and disappearing of those electric field vectors, coming from nowhere! And this happens with both objects. Therefore, the ad hoc use of creation and destruction operators \cite{awaya} begins also to make sense.\\
According to the model, self-interference is a natural consequence of the internal structure of the flat quantum particle, which because its higher dimensionality can be in two places at the same time and leaves electric field vectors with different phases in the world they are intersecting, and therefore could interfere with themselves. Also, the uncertainty principle appears as an intrinsic property of the quantum object, which is completely non-local (it is in two places at the same time) to provide information about its momentum and when it is localized in only one place, it stops to provide such information.\\
Self-interference and the uncertainty principle are central for the understanding of any interpretation of Quantum Mechanics (QM). For instance, those properties have been used as evidence for the parallel universes interpretation \cite{davies, shimony} However, the higher dimensionality of flat quantum objects, requires the use of just one universe (the plane), to explain quantum properties.\\
Since one has to deal with the intersection of higher dimensional object, which can be in two places at the same time, leaves a sinusoidal pattern and obeys the uncertainty principle. The Copenhagen interpretation, based on probabilities derived from the wave equation (the electric field sinusoidal pattern) and uses the principle of complementarity, appears as a good approximation. Albeit, the quantum object inner structure was abandoned by this interpretation. In this sense, the present model is consistent with that interpretation of QM. Therefore, it is highly probable for the electron and/or the photon to be in two places (both holes of the interference slit) at the same time. And this just reflexes a property of the intersection of a higher dimensional object in a lower dimensional space.
Finally, the new three dimensions described, help to understand why string theory proposes 10 to explain matter, light and gravity (everything). It would be possible that sub-atomic particles show others extra dimensions. However, with these new three, their quantum-optic behavior can be understood. Another interesting point is that these dimensions are curved, not at the Plank length (10$^{-33}$ m) as it has been established \cite{kaku}, but as it has been shown, it happens at atomic lengths (10$^{-15}$ m).
As a consequence of the number of phenomena that can be explained with the same model, \textbf{\textit{the intersection of a higher dimensional object in a space with lower dimensions has proven to be a powerful concept to explain quantum properties.}}

\section{Conclusions}
A new perspective to matter and light has been offered. In explaining self-interference a number of new dimensions and symmetries appears. Inertia and mass found its origin in the structure and kind of interaction with the space, done by the flat quantum object. The new concept of mass found with this model, joins quantum mechanics and the general theory of relativity. Nonetheless the dimensional constrain found by working with a lower than real dimensional model, consistency with the actual knowledge is found either by flat fermions and flat photon.
\section{Acknowledgement}
The author is deeply thankful to Dr. Humberto Figueroa, whose inspiration and discussions were crucial for this work and to Mr. Victor Blanco and Mr. Jos\'e Bergoya for the drawings.

 \end{document}